\documentclass[prb, fleqn,twocolumn, showpacs, showkeys]{revtex4}

\usepackage[dvips]{epsfig}

\begin{document}
\title{Incommensurate structure of the spin-Peierls compound TiOCl
}
\author{A. Krimmel$^1$, J. Strempfer$^{2}$, B. Bohnenbuck$^{2}$, B. Keimer$^{2}$, M. Hoinkis$^{3,4}$,
M. Klemm$^3$, S. Horn$^3$, A. Loidl$^1$, M. Sing$^4$, R.
Claessen$^4$, M. v. Zimmermann$^5$} \affiliation{ $^1$
Experimentalphysik V, Elektronische Korrelationen
und Magnetismus, Universit\"at Augsburg, D - 86159 Augsburg, Germany\\
$^2$ Max-Planck-Institut f\"ur Festk\"orperforschung,
Heisenbergstr. 1, D-70569 Stuttgart, Germany\\
$^3$ Experimentalphysik II, Institut f\"ur Physik, Universit\"at
Augsburg, D - 86159 Augsburg, Germany\\
$^4$ Experimentelle Physik 4, Universit\"at W\"urzburg, D-97074
Wuerzburg, Germany\\
$^5$ Hamburger Synhrotron Strahlungslabor (HASYLAB) at Deutsches
Elektronen-Synchrotron (DESY), Notkestr. 85, 22603 Hamburg,
Germany }

\email{Alexander.Krimmel@physik.uni-augsburg.de}

\date{\today}

\begin{abstract}
We report on a detailed single crystal x-ray diffraction study of
the unconventional spin-Peierls compound TiOCl. The intermediate
phase of TiOCl is characterized by an incommensurate modulation
which is virtually identical to that recently found in the
homologue compound TiOBr. The first order phase transition between
the spin-Peierls ground state and the incommensurate phase reveals
the same kind of thermal hysteresis in both, its crystal structure
and magnetic susceptibility. A weak, but significant magnetic
field effect is found for this phase transition with a field
induced shift of the transition temperature of $\Delta
T_{c1}=-0.13$~K for an applied field of $B=10$~T along the chain
direction. The field induced changes of the incommensurate crystal
structure are compatible with a scenario of competing intra- and
inter-chain interactions.
\end{abstract}



\maketitle

\section{\label{sec:intro}Introduction}
Strongly correlated low dimensional spin systems have attracted
great interest due to a large variety of fascinating physical
properties. The reduced dimensionality often allows for a
description by exactly solvable theoretical models which may
clarify fundamental quantum mechanical aspects of solids,
including phenomena of significant technological potential like
high-$T_c$ superconductivity. Moreover, the complex interplay
between the different microscopic degrees of freedom (charge,
spin, orbital and lattice degrees of freedom) is at the heart of
numerous phase transitions leading to exotic ground states.
Considering a one-dimensional (1D) antiferromagnetic (AFM) $S=1/2$
spin chain, a coupling to the lattice may result in a spin-Peierls
transition with a non-magnetic, dimerized ground state. The first
example of an inorganic spin-Peierls compound is CuGeO$_3$
\cite{Hase93}. More complex physics is obtained if the spins are
additionally coupled to charge or orbital degrees of freedom, as
manifested i. e. in a metal-to-insulator transition (MIT) in
Na$_{1/3}$V$_2$O$_5$ \cite{Heinrich04}.

Recently, the titanium-based oxohalides TiOX (X=Cl, Br) have been
discussed as new unconventional inorganic spin-Peierls systems
\cite{Seidel03,Imai03,Kataev03,Caimi04,Lemmens04,Hemberger05,Pisani05,Shaz05,
Hoinkis05,Ruckamp05,Kato05,Lemmens05,vansmaalen05}. They
crystallize in an orthorhombic structure with Ti-O bilayers within
the $ab$-plane well separated by Cl/Br ions \cite{Schafer58}.
Quasi-1D S=1/2 spin chains along the crystallographic $b$-axis are
formed via orbital ordering giving rise to strong direct exchange
with an exchange constant of $J/k_B \approx 660$~K
\cite{Seidel03}. For TiOCl, the low temperature spin-Peierls state
is established by a steep decrease of the magnetic susceptibility
below $T_{c1}=67$~K \cite{Seidel03}, accompanied by a simultaneous
lattice dimerization of the Ti$^{3+}$ ions, as evidenced by
corresponding superlattice reflections showing a doubling of the
unit cell along the $b$-axis \cite{Shaz05}. Infrared
\cite{Caimi04} and Raman spectroscopy \cite{Lemmens04}, as well as
electron spin resonance (ESR) \cite{Kataev03} and NMR experiments
\cite{Imai03} corroborated these results and revealed a spin
excitation gap of $\Delta = 430$~K. However, a conventional
spin-Peierls scenario is insufficient to account for the physical
behaviour of TiOCl. A wealth of experimental results have
established a first order phase transition from the spin-Peierls
ground state into a second, intermediate phase at $T_{c1}=67$~K,
which extends up to $T_{c2}=91$~K where a second order phase
transition separates the intermediate phase from the normal
paramagnetic state at high temperatures. The nature of the
intermediate phase of TiOCl has not yet been clarified. Based on
the temperature dependence of the $g$-factors and line width in
ESR experiments \cite{Kataev03} and, in particular, on phonon
anomalies found in Raman and IR spectroscopy
\cite{Caimi04,Lemmens04} it has been proposed that orbital
fluctuations play an essential role and may extend well above
$T_{c2}$ up to 130 K. This interpretation has been further
supported by recent specific heat measurements \cite{Hemberger05}
and is also in agreement with electronic structure calculations
\cite{Seidel03,Saha04}. Within density functional theory employing
the LDA+U approximation, the electronic ground state configuration
$3d^1_{xy}$ of the Ti$^{3+}$ ions can couple to optical phonon
modes that in turn may lead to strong orbital fluctuations within
the $t_{2g}$ crystal field multiplet \cite{Seidel03,Saha04}.
Moreover, the importance of correlation effects has been revealed
by combined LDA+DMFT studies \cite{Saha05,Hoinkis05}.

In contrast, ARPES measurements could not detect any evidence for
phonon assisted orbital fluctuations \cite{Hoinkis05}. Moreover,
recent polarization dependent optical measurements in combination
with cluster calculations provide evidence that the orbital
degrees of freedom are actually quenched \cite{Ruckamp05}.
Alternatively, it has been proposed that inter-chain interactions
within the bilayers lead to an incommensurate spin-Peierls state
below $T_{c2}$ that locks-in in a conventional commensurate
dimerized phase below $T_{c1}$ \cite{Ruckamp05}.

In fact, such a behaviour has been observed in the homologue
compound TiOBr \cite{vansmaalen05}. TiOBr and TiOCl have the same
electron configuration and crystal structure and exhibit similar
physical properties. Like TiOCl, TiOBr shows two phase transitions
at $T_{c1}=27$~K and $T_{c2}=47$~K. The low temperature phase of
TiOBr also shows a twofold superstructure along the $b$-axis in
agreement with a spin-Peierls ground state
\cite{sasaki05,vansmaalen05}. Recently, a single crystal x-ray
diffraction study on TiOBr revealed an incommensurate modulated
structure for the intermediate phase \cite{vansmaalen05}, in
accordance with optical data and cluster calculations.

Here we report on a detailed x-ray diffraction study on single
crystalline TiOCl to elucidate the nature of the intermediate
phase and to investigate any magnetic field effect of the phase
transitions. This is motivated by a large magnetic field
dependence of the incommensurate structure observed above a
threshold magnetic field in other spin-Peierls compounds such as
TTF-CuBDT \cite{Kiryukhin95} and CuGeO$_3$ \cite{Kiryukhin96}.

\section{\label{sect:expres}Experimental results}

Single crystals of TiOCl were prepared by chemical vapour
transport \cite{Schafer58} from the starting materials TiCl$_3$
and TiO$_2$. The samples were characterized by magnetization
measurements employing a SQUID magnetometer and the magnetic
properties were found in excellent agreement with published
results. Magnetic field dependent single crystal x-ray diffraction
measurements employing synchrotron radiation have been performed
at the beamline BW5 of HASYLAB (DESY, Hamburg). An incident photon
energy of 100 keV was used. The sample was mounted in a cryomagnet
allowing for temperatures $1.6 \le T \le 300$~K in horizontal
fields up to $B=10$~T. The sample with a size of 1x1x0.01~mm$^3$
was oriented with the $bc$-plane in the horizontal scattering
plane. The magnetic field was oriented along the scattering vector
in the chain direction. This geometry was possible due to the
small scattering angles at high photon energies. By tilting the
cryomagnet, also small values in $h$ were accessible. At low
temperatures ($T=10$~K), a number of superlattice reflections
along the chain direction $(0, k+0.5, 0), k=0, 1, 2$ have been
recorded. The strongest intensity was found for $(0, 1.5, 0)$.

\begin{centering}
\begin{figure}[h!]
\epsfig{file=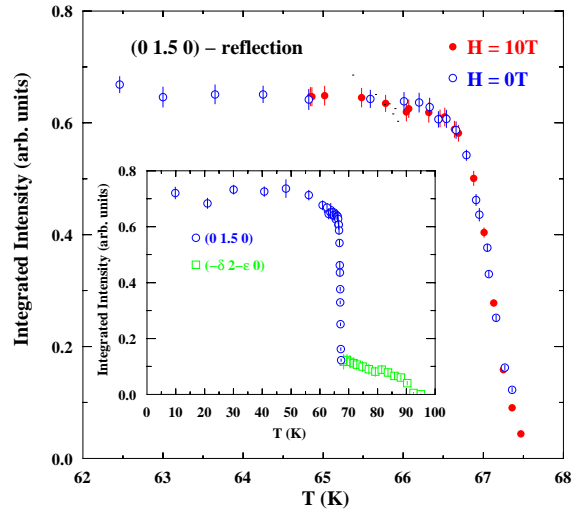,angle=0,width=0.44\textwidth}
\caption{(Color online) Temperature dependence of the intensity of
the (0, 1.5, 0) superlattice reflection of TiOCl in the vicinity
of the first phase transition around $T_{c1}=67$~K. Shown are
measurements in zero field (open circles) and in an external field
of $B=10$~T (full circles), respectively. The inset shows
intensities of both commensurate and incommensurate reflections
over the whole temperature range.} \label{TdepInt}
\end{figure}
\end{centering}

Fig.~\ref{TdepInt} shows the temperature dependence of the
intensity of the $(0, 1.5, 0)$ reflection around the first phase
transition at $T_{c1}=67.5$~K in zero field and in an external
field of $B=10$~T. As evident from Fig.~\ref{TdepInt}, the
intensity remains virtually constant at low temperatures, starts
to steeply decrease at 66.5 K and vanishes at 67.5 K. Within the
experimental accuracy, no differences between the measurements in
zero field and $B=10$~T are observed with a marginal possible
error of the temperature below 0.01 K. Cycling the temperature
reveals a pronounced thermal hysteresis of the intensity of the
$(0, 1.5, 0)$ reflection.

\begin{centering}
\begin{figure}[h!]
\epsfig{file=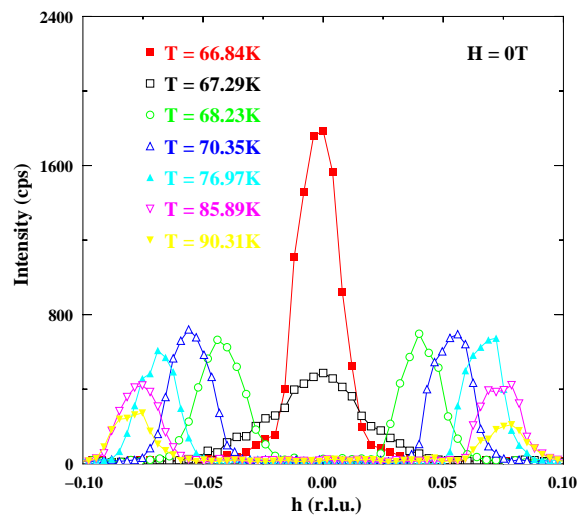,clip,angle=0,width=0.44\textwidth}
\caption{(Color online) Evolution of the (0, 1.5, 0) superlattice
reflection of TiOCl for various temperatures on passing through
the first phase transition from the dimerized spin-Peierls ground
state into the intermediate phase in zero field. The peak
splitting indicates an incommensurate modulation of the
intermediate phase.} \label{Tdepsplit}
\end{figure}
\end{centering}

Fig.~\ref{Tdepsplit} shows the evolution of the low temperature
$(0, 1.5, 0)$ reflection for increasing temperatures, covering
essentially the temperature range of the second, intermediate
phase 67.5 K $=T_{c1} \le T \le T_{c2}=92.5$~K. At $T_{c1}$ the
$(0, 1.5, 0)$ reflection splits into two incommensurate satellites
which also show an additional incommensurate component along $k$.

\begin{figure}[h!]
\epsfig{file=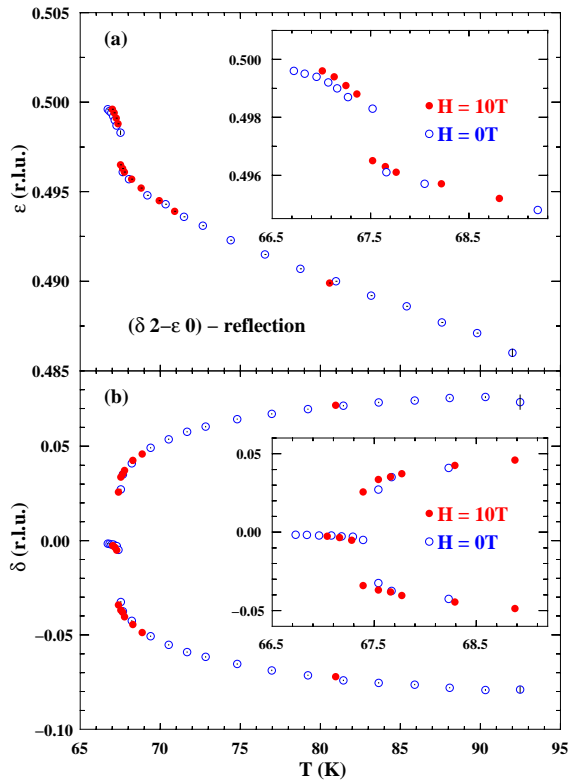,width=0.44\textwidth} \caption{(Color
online) Temperature dependence of the incommensurate components
$\epsilon$ (a) and $\delta$ (b) of the $(\pm\delta, 2 - \epsilon,
0)$ reflection of TiOCl in the intermediate phase in zero field
(open circles) and in an external field of $B=10$~T (full
circles). The insets show the incommensurate positions $\delta$
and $\epsilon$ in more detail around $T_{c1}$.} \label{TdepPos}
\end{figure}

The observed incommensurate reflections are $(\pm \delta,
\epsilon, 0), (\pm \delta, 2-\epsilon, 0)$ and $(\pm \delta,
2+\epsilon, 0)$ consistent with space group $Pmmn$. The
incommensurately modulated intermediate phase can be described by
a propagation vector ${\bf q}=(\pm \delta, 0.5+\epsilon, 0)$ with
$0 \le \delta \le 0.078$ and $0.4857 \le \epsilon \le 0.5$. The
satellites can be monitored up to $T=92.5$~K where the undistorted
orthorhombic structure of the paramagnetic phase is recovered. The
temperature dependence of the two incommensurate components of the
reflection $(\pm \delta, 1.5+\epsilon, 0)$ is shown in
Fig.~\ref{TdepPos}. Fig.~\ref{TdepPos}a, shows the $k$-component
$\epsilon$ and Fig.~\ref{TdepPos}b, the $h$-component $\delta$ of
the incommensurate satellite for zero field and in an applied
field of $B=10$~T. An expanded view around $T_{c1}$ is given in
the corresponding insets. For both components, a small, but
significant shift $\Delta T_{c1}= -0.13$~K of the phase transition
temperature is observed in the external field of 10 T along the
chain direction. Remarkably, the transition temperatures for the
modulations along the $h$- and $k$-direction appear to be slightly
different. A scan along $k$ at $T=67.52$~K shows a single peak at
$k=1.5015$ whereas a scan along $h$ shows a peak splitting with
$\delta= \pm 0.03$ at $T=67.54$~K. Moreover, also the intensities
of the satellite reflections exhibit slight changes thus
confirming a small field induced modification of the
incommensurate crystal structure.

\section{\label{sect:conc}Discussion and Conclusion}

We have performed a detailed single crystal x-ray diffraction
study of the spin-Peierls compound TiOCl. However, it should be
noted that the present investigation does not represent a complete
crystal structure determination. Due to geometric restrictions by
use of a large cryomagnet, only reflections of type $(0, k, 0)$
were explored in detail for their temperature dependence and
possible field effects. Therefore, the focus was on the phase
transitions in order to elucidate the nature of the intermediate
phase of TiOCl and its relation to the unconventional spin-Peierls
transition.

The temperature dependence of the $(0, 1.5, 0)$ reflection is
characteristic for a doubling of the unit cell along the $b$-axis
due to the dimerized spin-Peierls ground state \cite{Shaz05}. A
sudden decrease of the intensity is observed at $T_{c1}=67.5$~K
which confirms that the transition is of first order. Moreover, a
corresponding thermal hysteresis of the intensity of this
superlattice reflection is observed. In combination with the same
type of hysteresis found in the magnetic susceptibility, it is
concluded that the non-magnetic, dimerized ground state of TiOCl
is realized via a first order spin-Peierls transition.

Recently, the unconventional properties of TiOX (X=Cl, Br) have
been interpreted in terms of frustrated interchain interactions
within the bilayers \cite{Ruckamp05}. Within this scenario, the
spin-Peierls mechanism would give rise to an intermediate phase
characterized by an incommensurate order with a subsequent lock-in
transition into the commensurate dimerized ground state
\cite{Ruckamp05}. In fact, such an incommensurate modulated
structure has recently been observed in TiOBr
\cite{vansmaalen05}. The intermediate phase of TiOCl is also
characterized by such an incommensurate structural modulation with
a propagation vector ${\bf q}=(\pm \delta, 0.5+\epsilon, 0)$ with
$0 \le \delta \le 0.078$ and $0.4857 \le \epsilon \le 0.5$, as
evidenced by a corresponding peak splitting. The intermediate
phase of TiOCl exhibits a two dimensional (2D) modulation within
the Ti-O bilayers. The absolute values and the temperature
dependence of the modulation vector in TiOCl are almost identical
to those observed for TiOBr (for TiOBr, the published temperature
dependence of the modulation vector is restricted to the
$x$-component or $\delta$) \cite{vansmaalen05}. We therefore
conclude that both, TiOCl and TiOBr exhibit the same kind of
incommensurate modulation in their intermediate phase. Apart from
the incommensurability, this modulation is characterized by rather
large displacements along the $b$-axis (chain direction) and
comparable small amplitudes along the $a$-axis
\cite{vansmaalen05}. The intermediate phase can be interpreted as
either due to frustrated spin-Peierls interactions
\cite{Ruckamp05} or, alternatively, as a phase with competing 1D
spin-Peierls interactions and 2D magnetic interactions which are
coupled to the lattice modulation \cite{vansmaalen05}.

A crucial test to identify a spin-Peierls state is its generic
$B-T$-phase diagram \cite{Cross79}. Due to the large energy scale
in TiOCl with an exchange constant of $J/k_B \approx 660$~K no
significant magnetic field effects are expected within the
accessible field range of conventional laboratory magnets. Our
measurements in an external field of $B=10$~T could not observe
any significant change of the principal superlattice reflections
of type $(0, k+0.5, 0)$ characterizing the doubling of the
$b$-axis due to the commensurately dimerized spin-Peierls ground
state. However, a weak but significant field effect is found for
the phase transition into the incommensurate phase with a field
induced shift of the transition temperature of $\Delta
T_{c1}=-0.13$~K for $B=10$~T. The negative temperature shift
indicates a stabilization of the incommensurate structure and a
suppression of the antiferromagnetic spin singlet formation of the
spin-Peierls ground state by the external field. These
observations may be accounted for within the scenario of
frustrated inter-chain interactions that give rise to a second,
incommensurate phase \cite{Ruckamp05}. In TiOCl, the formation of
spin chains along the $b$-axis with a strong antiferromagnetic
exchange results from orbital ordering of the $xy$-orbitals
\cite{Seidel03,Saha04} with a corresponding strong magneto-elastic
coupling. The orbital ordering further gives rise to inter-chain
$\pi$-bonding with a predominantly ferromagnetic exchange. Based
on band structure calculations \cite{Saha04}, the inter-chain
exchange can be estimated to be one order of magnitude weaker than
the principal exchange along the chains. The competition between
these different exchange interactions leads to frustration and an
incommensurately modulated crystal structure \cite{Ruckamp05}. The
application of an external magnetic field then slightly shifts the
equilibrium position of these competing interactions. The small
magnetic field effect on the incommensurability confirms that the
incommensurate modulation is of fundamentally different origin as
the field-induced modulation observed in classical spin-Peierls
systems \cite{Kiryukhin95,Kiryukhin96}. To conclude, we have
determined the incommensurate modulation of the intermediate phase
of TiOCl and find some small, but significant changes upon the
application of an external field of $B=10$~T along the chain
direction. These results are compatible with a frustrated
spin-Peierls scenario.

\acknowledgments

This work was supported by the German Bundesministerium f\"ur
Bildung und Forschung under contract No. VDI/EKM 13N6917 and by
the Deutsche Forschungsgemeinschaft DFG via SFB 484, Augsburg.

\end{document}